\documentclass[11pt]{article}
\usepackage{amsmath}
\usepackage{amssymb}
\usepackage{latexsym}
\usepackage{epsfig}

\newcommand{\be}{\begin{equation}}
\newcommand{\ee}{\end{equation}}

\begin{document}

\vskip 3.0cm

\centerline{\Large \bf One-dimensional quantum channel and Hawking radiation }
\vspace*{2.0ex}
\centerline{\Large \bf of the Kerr and Kerr-Newman black holes}
\vspace*{10.0ex}

\centerline{\large Deyou Chen \footnote{Email: ruijianchen@gmail.com}$^{ab}$, Benrong Mu
\footnote{Email: mubenrong@uestc.edu.cn}$^{a}$, Houwen Wu \footnote{Email:
iverwu@uestc.edu.cn}$^{a}$, Haitang Yang \footnote{Email:
hyanga@uestc.edu.cn}$^{a}$}

\vspace*{7.0ex}

\vspace*{4.0ex}

\centerline{\large \it  $^{a}$School of Physical Electronics,}

\centerline{\large \it
University of Electronic Science and Technology of China,}

\centerline{\large \it Chengdu, 610054, China.}
\vspace*{2.0ex}

\centerline{\large \it and}
\vspace*{2.0ex}

\centerline{\large \it  $^{b}$Institute of theoretical physics,}

\centerline{\large \it
China West Normal University,}

\centerline{\large \it Nanchong, 637009, China.}

\vspace*{10.0ex}

\centerline{\bf Abstract}
\bigskip
\smallskip

In this paper, we review the one-dimensional quantum
channel and investigate Hawking radiation of bosons and fermions in
Kerr and Kerr-Newman black holes. The result shows the Hawking radiation
can be described by the quantum channel. The thermal conductances are
derived and related to the black holes' temperatures.

\vfill \eject

\baselineskip=16pt

\vspace*{10.0ex}

\tableofcontents

\section{Introduction}

Black holes are not black and radiate thermally \cite{Hawking},
which reveals the relation between quantum theory and gravity
theory. When first proved it, Hawking described the radiation as a
tunnelling process caused by vacuum fluctuations near the black hole
horizon. Where a virtual particle pair spontaneously creates near
the horizon. The negative energy particle is absorbed by the black
hole, while the positive energy particle is left outside the
horizon, moves to infinite distance and forms Hawking radiation.

There are several methods to derive Hawking radiation. Hawking's
original derivation is directly physical and dependent on the
calculation of Bogoliubov coefficient \cite{Hawking}. The black
hole's background spacetime was seen as fixed one with the emission
of particles. In the method of Damour, Ruffini and Sannan
\cite{DR,SS}, the second quantizen is avoided and Hawking radiation
is derived by relativistic quantum mechanics in the curved
spacetime. There is no need to demand the thermal equilibrium between
black holes and the environment outside the black holes and consider
black hole's collapse. The radiation spectrum was obtained and given
by the Planck distribution as \cite{Hawking,DR,SS}

\begin{equation}
N^\pm \left( \omega \right) = \frac{1}{e^{\frac{\omega}{k_B T_H}}\pm
1}, \label{eq:1}
\end{equation}

\noindent where + (-) correspond to the fermion (boson), $k_B$ and
$\omega $ are the Boltzmann constant and the energy of emission
particle, respectively. The semi-classical tunnelling method, based
on particles in a dynamical geometry, was
put forward by Parikh and Wilczek \cite{KWP}. Their work shown that
the radiation spectrum is not purely thermal one and the tunnelling
rate is related to the change of Bekenstein-Hawking entropy. This
result gives the leading correction to Hawking radiation spectrum.
The method based on the calculation of Hawking fluxes has a long
history \cite{WU,BMPS,TP,BZ,CFV,RW,IUW1}. The expectation value of
energy-momentum tensor in the vacuum state was obtained and connects
the black hole's temperature. In \cite{WU}, the expectation value
with respect to the Unruh vacuum was derived, which gave a elegant
description for the radiation. In \cite{RW,IUW1}, the fluxes of
Hawking radiation were derived by the cancellation of gravity
anomaly and gauge anomaly at the horizon. It shown the scalar field
theory of a arbitrary $(D+2)$-dimensional black hole can be reduced
to $(1+1)$-dimensional quantum field theory by a dimensional
reduction technology.

On the other hand, the holographic principle \cite{Maldacena,
Susskind} shows that a generic physical system in three-dimensions
(3D) can be seen as two-dimensional (2D). Considering this case,
Bekenstein and Mayo \cite{BM} gave a further constriction of
dimensions and shown the black hole in 3D behaves as the
one-dimensional (1D) channel with the consideration of entropy
or information flow. In this work, the entropy flux was researched.
The maximum entropy rate was obtained and is the same as that of
Pendry's result \cite{JBP}.

The 1D quantum channel was first put forward in \cite{IL} and
applied to the calculation of conductance. In \cite{IL}, the authors
described the channel as follows: an ideal channel adiabatically
connects two reservoirs. The reservoirs are the electronic
equivalent of a radiative blackbody; the electrons coming out of a
reservoir are occupied according to the Fermi distribution that
characterizes the deep interior of that reservoir. If the channel is
narrow enough, only the lowest of the transverse eigenstates in the
channel has its energy below the Fermi level. Then the channel can
be effectively seen as 1D. Subsequently, people introduced the
channel to research on the fluxes of entropy and energy and others
topics \cite{RK,KM,BV,SUNBRB}.

Very recently, the energy flux and entropy flux of Hawking radiation
of photons in a Schwarzschild spacetime has been seen as 1D
Landauer transport process \cite{NBN}. The energy flux of photons
in a individual single-channel was obtained and is identical to the
outgoing Hawking flux. Then the authors concluded that Hawking
radiation process of photons can be described by a 1D quantum
channel and the channel connects two thermal baths with one side
being the black hole and another side being the outside thermal
environment surrounding the black hole. In their work, the and
entropy flux and net entropy production in $(1+1)$-dimensions were
also investigated in detail. The derivation of Hawking flux was
dependent on the expectation value for the stress-energy tensor in
the conformal structure. The expectation value with respect to the
Unruh vacuum \cite{WU} was derived and the flux seen by an inertial observer
at infinity distance was only related to the black hole's temperature.

Our aim in this paper is to use the Landauer transport model to
investigate Hawking radiation of bosons and fermions in Kerr and
Kerr-Newman black holes. In the Kerr and Kerr-Newman black holes,
Hawking fluxes contains the energy-momentum tensor flux, the
angular momentum flux and the charge flux(for the charged particle
in the Kerr-Newman spacetime). Meanwhile, due to the existence of
dragging effect of coordinate system in the rotating spacetime,
the matter field in the ergosphere near the horizon is dragged by
the gravitational field with an azimuthal angular velocity,
therefore the chemical potential should describes this dragging
effect. Meanwhile, in the Kerr-Newman black hole due to the effect
of electromagnetic field, the chemical potential should also
contain electromagnetic potential. The magnetic quantum number
of the particle is regarded as a general charge in this paper. Both
of Hawking radiation of the boson and the fermion are described by a
1D quantum channel in this paper. The thermal conductances of the
Kerr and Kerr-Newman black holes are derived by the 1D channel
description and related to the Hawking temperatures.

The rest of this paper is outlined as follows. In sect. 2, from the
formulae of charge and energy fluxes in the 1D quantum channel,
we get the expressions of charge flux and energy flux of a boson
and a fermion. In sect. 3, Hawking radiation of the boson and the
fermion in the Kerr black hole is obtained and described by the 1D
channel. Extending this work to charged rotating black holes, we
investigate the Hawking radiation of charged particles in a
Kerr-Newman spacetime in sect. 4, Sect. 5 contains some discussions and conclusions.

\section{Review the one-dimensional quantum channel}

The 1D quantum channel was first put forward in \cite{IL}.
Subsequently, Rego and Kirczenow gave the formulae of energy flux
and charge flux in the channel \cite{RK}. One can refer to
\cite{IL,RK,KM,BV,SUNBRB,NBN} in detail. In this section, we
review the 1D quantum channel. In a two terminal transport
experiment two infinite reservoirs are adiabatically connected
to each other by a 1D channel. $T$ and $\mu $, which are independent
variables, denote temperatures and chemical potentials of the
reservoirs, respectively. In case of a reservoir with charged
particles, $\mu $ is the electrochemical potential energy and is the
combination of chemical potential and electrostatic particle
energy governed by the external field. In a signal channel, the
currents of charge and energy flowing from the lift (L) and right
(R) reservoirs are given by

\begin{equation}
I_{R\left( L \right)} = \frac{q k_B T_{R\left( L \right)} }{2\pi
\hbar }\int\limits_{ - \frac{\mu }{k_B T_{R\left( L \right)}
}}^\infty f \left( x,g \right)dx,\label{eq:2}
\end{equation}

\begin{equation}
\dot {E}_{R\left( L \right)} = \frac{\left( {k_B T_{R\left( L
\right)} } \right)^2}{2\pi \hbar }\int\limits_{ - \frac{\mu }{k_B
T_{R\left( L \right)} }}^\infty {f \left( x,g \right)\left( {x +
\frac{\mu }{k_B T_{R\left( L \right)} }} \right)dx} ,\label{eq:3}
\end{equation}

\noindent with the fractional exclusion statistics $f(x,g)$. In this
paper, we focus our attention on identical particles system. The
distribution function was derived by Wu \cite{YSW} and for an idea
gas of particles it obeys the fractional exclusion statistics
$f\left( x,g \right) = \left( {\omega \left[ {\frac{x - \mu }{k_B
T}} \right] + g} \right)^{ - 1}$. The relation between $x$ and $w$
is $\omega ^g\left( {1 + \omega } \right)^{1 - g} = e^x$, where $g$
denotes the statistical interaction and describes bosons for $g = 0$
and fermions for $g = 1$.

For a boson, the formulae of charge current and energy
current are expressed as

\begin{equation}
I_{R\left( L \right)} = \frac{q k_B T_{R\left( L \right)} }{2\pi
\hbar }\int\limits_{ - \frac{\mu }{k_B T_{R\left( L \right)}
}}^\infty {\frac{1}{e^x - 1}dx} ,\label{eq:4}
\end{equation}

\begin{equation}
\dot {E}_{R\left( L \right)} = \frac{\left( {k_B T_{R\left( L
\right)} } \right)^2}{2\pi \hbar }\int\limits_{ - \frac{\mu }{k_B
T_{R\left( L \right)} }}^\infty {\frac{1}{e^x - 1}\left( {x +
\frac{\mu }{k_B T_{R\left( L \right)} }} \right)dx} . \label{eq:5}
\end{equation}

\noindent Due to the particularity of the boson, it is difficult to
get values of the above equations. However, photon is a especial
case and has no the rest mass and chemical potential. So there is
only the energy current for the photon, which is derived as

\begin{equation}
\dot {E}_{R\left( L \right)}  = \frac{\pi k_B^2 T_H^2 }{12\hbar }.
\label{eq:6}
\end{equation}

For a fermion, the formulae of charge current and energy
current are written as

\begin{equation}
I_{R\left( L \right)} = \frac{q k_B T_{R\left( L \right)} }{2\pi
\hbar }\int\limits_{ - \frac{\mu }{k_B T_{R\left( L \right)}
}}^\infty {\frac{1}{e^x + 1}dx}  = \frac{q\mu _{R\left( L \right)}
}{2\pi \hbar }, \label{eq:7}
\end{equation}

\begin{equation}
\dot {E}_{R\left( L \right)} = \frac{\left( {k_B T_{R\left( L
\right)} } \right)^2}{2\pi \hbar }\int\limits_{ - \frac{\mu }{k_B
T_{R\left( L \right)} }}^\infty {\frac{1}{e^x + 1}\left( {x +
\frac{\mu }{k_B T_{R\left( L \right)} }} \right)dx}
= \frac{\mu _{R\left( L \right)}^2 }{4\pi \hbar } + \frac{\pi k_B^2 T_H^2
}{12\hbar }. \label{eq:8}
\end{equation}

\noindent Now we give some physical expatiation on the above
equations. Due to the particularity of the boson system, it is
difficult to derive the value of energy current and charge current.
However, photons are most simple case, so the energy current is
easily obtained. For the fermion, Eq. (\ref{eq:6}) shows that the
charge current is related to the charge and the corresponding
chemical potential of transmission particles. Eq. (\ref{eq:7}) tells
us that the energy flux flowing through the 1D system contains two
independent components: the first component is the flux of
particles, which is determined by the chemical potential $(\mu )$;
the second component is only related to the temperature $(T)$ of the
reservoirs. In the following, we use the 1D channel to describe
Hawking radiation of bosons and fermions in the curved spacetimes.

\section{Hawking radiation in the Kerr spacetime}

The thermodynamic property of black holes is an important topic and
attracts much attention. The reason is that the research on black
holes concerns quantum theory and gravity theory, while the gravity
theory of quantum has not been solved commendably now. Therefore it
helpful to solve this theory by the research on black holes. In this
section, we investigate Hawking radiation of a boson and a fermion
in the Kerr background spacetime \cite{Kerr}. There are several
derivations of Hawking radiation. One derivation is focused on the
investigation of the Hawking fluxes, which can be obtained from the
expectation value of energy-momentum tensor in the Unruh vacuum or
from the cancellation of anomaly. In this paper, the Hawking flux is
directly obtained from the radiation spectrum. The thermal spectrum
in the Schwarzschild background spacetime was gotten in
eq.(\ref{eq:1}) with the Hawking temperature $T_H$. For the
$4D$ Kerr spacetime, which is given by

\begin{equation}
ds^2 = - \frac{\Delta }{\rho ^2}\left( {dt - a\sin ^2\theta d\phi }
\right)^2 + \frac{\rho ^2}{\Delta }\left( {dr^2 + \Delta d\theta ^2}
\right) + \frac{\sin ^2\theta }{\rho ^2}\left[ {adt - \left( {r^2 +
a^2} \right)d\phi } \right]^2 , \label{eq:9}
\end{equation}

\noindent with $\rho ^2 = r^2 + a^2\cos ^2\theta $, $\Delta = r^2 -
2Mr + a^2 = \left( {r - r_H } \right)\left( {r - r_ - } \right)$,
the black hole mass $M$ and the angular momentum unit mass $a$, the
thermal spectrum at the Hawking temperature is derived by

\begin{equation}
N_m^\pm \left(\omega \right) = \frac{1}{e^{\frac{\omega -
m\Omega_H}{k_B T_H}}\pm 1}, \label{eq:10}
\end{equation}

\noindent where + (-) correspond to the fermion and boson, $\omega$
and $ m $ are the energy and the magnetic quantum number of the
emission particle, respectively. One can refer to \cite{LX} for
Hawking radiation of the fermion. $\Omega_H $ and $T_H$ are
respectively the angular velocity and the Hawking temperature at the
outer horizon and are derived as

\begin{equation}
\Omega_H=\frac{a}{r_H^2 + a^2}, \quad T_H = \frac{r_H - r_-}{4\pi
(r_H^2 + a^2)}, \label{eq:11}
\end{equation}

\noindent with the outer (inner) horizons $r_H = M + \sqrt {M^2 -
a^2} \quad (r_ - = M - \sqrt {M^2 - a^2})$. We first investigate the
Hawking fluxes of the boson. Due to the axisymmetrical property, the
Hawking fluxes contains two parts: angular momentum flux and
energy-momentum tensor flux. Using eqs. (\ref{eq:10}) and
(\ref{eq:11}), we derive the fluxes of angular momentum and
energy-momentum tensor as

\begin{equation}
F_a = m\int\limits_0^\infty {\frac{N_m^ - \left( \omega
\right)}{2\pi \hbar }d\omega } = \frac{m}{2\pi \hbar
}\int\limits_0^\infty {\frac{1}{e^{\frac{\omega - m\Omega _H }{k_B
T_H }} - 1}d\omega }=\frac{m k_B T_{H} }{2\pi \hbar }\int\limits_{ - \frac{ m\Omega
_H}{k_B T_{H} }}^\infty {\frac{1}{e^x - 1}dx} , \label{eq:12}
\end{equation}

\begin{equation}
F_M = \int\limits_0^\infty {\frac{N_m^- \left( \omega \right)}{2\pi
\hbar }\omega d\omega } = \frac{1}{2\pi \hbar }\int\limits_0^\infty
{\frac{\omega }{e^{\frac{\omega - m\Omega _H }{k_B T_H }} -
1}d\omega }= \frac{\left( {k_B T_{H} } \right)^2}{2\pi \hbar }\int\limits_{ -
\frac{ m\Omega _H }{k_B T_{H} }}^\infty {\frac{1}{e^x - 1}\left( {x
+ \frac{ m\Omega _H }{k_B T_{H} }} \right)dx} . \label{eq:13}
\end{equation}

\noindent The last equal signs in eqs. (\ref{eq:12}) and
(\ref{eq:13}) are obtained with a transformation $x = \frac{\omega - m\Omega
_H}{k_{B}T_{H}}$ . In \cite{IUW2}, the Hawking
fluxes in the scalar field were derived by cancellation of gravity
anomaly and gauge anomaly. Here the expression of the Hawking fluxes
are gotten, but it is difficult to get the value of the integral.
The reason is the particularity of the bosonic system. However,
it does not affect our final result. In the case of a photon, its
rest mass and the magnetic quantum number are zero. Therefore the
angular momentum flux (\ref{eq:12}) is zero and the energy-momentum
tensor flux (\ref{eq:13}) is reduced to

\begin{equation}
F_M = \frac{\pi k_B^2 T_H^2 }{12\hbar }. \label{eq:14}
\end{equation}

In the following, we investigate the Hawking fluxes of the fermion.
In \cite{RW}, the authors shown that the flux for thermal radiation
of massless bosons is $2$ times that of massless fermions. The same
result was also found in \cite{DU}. Subsequently this phenomena is
explained as that a single massless bosonic field is equivalent to
the fermionic field with a massless particle plus its antiparticle
\cite{Davies}. Therefore the Hawking fluxes of fermions should
contain the contributions of fermions and anti-fermions. For the fermion,
using eqs. (\ref{eq:10}) and (\ref{eq:11}), the total angular momentum
flux and the energy-momentum tensor flux are gotten as

\begin{equation}
F_a = m\int\limits_0^\infty {\frac{\left( {N_m^ + \left( \omega
\right) + N_{ - m}^ + \left( \omega \right)} \right)}{2\pi \hbar }}
d\omega= \frac{m}{2\pi \hbar }\int\limits_0^\infty {\left(
{\frac{1}{e^{\frac{\omega - m\Omega _H }{k_B T_H }} + 1} +
\frac{1}{e^{\frac{\omega + m\Omega _H }{k_B T_H }} + 1}}
\right)d\omega } = \frac{m^2\Omega _H }{2\pi \hbar }, \label{eq:15}
\end{equation}

\begin{equation}
F_M = \int\limits_0^\infty {\frac{\left( {N_m^ + \left( \omega
\right) + N_{ - m}^ + \left( \omega \right)} \right)}{2\pi \hbar }}
\omega d\omega
= \frac{1}{2\pi \hbar }\int\limits_0^\infty {\left(
{\frac{1}{e^{\frac{\omega - m\Omega _H }{k_B T_H }} + 1} +
\frac{1}{e^{\frac{\omega + m\Omega _H }{k_B T_H }} + 1}}
\right)\omega d\omega } = \frac{m^2\Omega _H^2 }{4\pi \hbar } +
\frac{\pi k_B^2 T_H^2 }{12\hbar }. \label{eq:16}
\end{equation}

\noindent Hawking radiation of the photon in the Schwarzschild
spacetime was described by the 1D quantum channel in \cite{NBN}. In
this section, we introduce this view to investigate Hawking
radiation of the boson and the fermion. The related parameters are
given as follows. The black hole temperature $T_H$ was derived in
eq. (\ref{eq:11}) and the thermal environment surrounding the black
hole is seen as $T_E = 0$ \cite{NBN}. We know the chemical potential
equals that the charge multiplies the corresponding potential. For
the Kerr black hole, due to the dragging effect of the coordinate
system, the matter field in the ergosphere near the horizon must be
dragged by the gravitational field with an azimuthal angular
velocity. Therefore the chemical potential of the particle in
this spacetime should contains this effect. In this paper, the
magnetic quantum number of the particle is regarded as a
general/gauge charge. So its chemical potential equals that the magnetic
quantum number multiplies the corresponding angular velocity, namely
$\mu = m\Omega_H$. While its chemical potential $\mu_E$ is regarded
as zero in the outside environment surrounding the black hole.

Now we return to the 1D channel. For the fermion, the total charge
current and energy current between the two reservoirs obtained from
eqs. (\ref{eq:7}) and (\ref{eq:8}) are

\begin{equation}
F = I_R - I_L = \frac{q}{2\pi \hbar} (\mu_R -\mu_L ), \label{eq:17}
\end{equation}

\begin{equation}
\dot {E} = \dot {E}_R - \dot {E}_L = \frac{1}{4\pi \hbar}(\mu_R ^2 -
\mu_L ^2) + \frac{\pi k_B^2}{12\hbar }(T_R^2 - T_L^2 ),
\label{eq:18}
\end{equation}

\noindent which are respectively identical to the angular momentum
flux (\ref{eq:16}) and the energy-momentum tensor flux (\ref{eq:17})
under the condition that $T_R = T_H$, $T_L = T_E $, $\mu_R = \mu $ and $\mu_L = \mu_E$.
For the boson, we can also get that the total charge current and
energy current obtained from (\ref{eq:5}) and (\ref{eq:6}) are
respectively identical to the angular momentum flux (\ref{eq:12})
and the energy-momentum tensor flux (\ref{eq:13}) under the same condition.
Photon is a sort of bosons. For a photon, both of the chemical potentials are zero in
the black hole and the outside environment surrounding the black
hole. The case of the photon has been contained in that of the
boson, so we do not investigate it again. Our result shows the
Hawking radiation of fermions and bosons in the Kerr spacetime can
be described by a 1D quantum channel, which is in consistence with
that of Nation \cite{NBN}, meanwhile gives the proof of the view of
Bekenstein and Mayo \cite{BM}.

Considering an observer is in the dragging field, he/she must move
with the dragging field. Then he/she can not feel the flow of
angular momentum flux and there is only the flow of the
energy-momentum flux for his/her feel. We can also prove the energy
current is identical to the energy-momentum tensor flux. The Hawking
fluxes in the dragging field have been studied by the cancellation
of anomaly in \cite{JW}. In this paper, we do not investigate this
case.

If the black hole is seen as 1D \cite{BM}, we can use parameters
of the 1D channel to describe that of the black hole. In \cite{RK},
the 1D thermal conductance was derived as $\kappa = \frac{\pi k_B^2
T}{6\hbar }$, which is only related to the temperature. Replacing
the temperature $T $ with the Hawking temperature $ T_H$, we get the
thermal conductance of a single channel in the Kerr black hole. It
should be emphasized here. In the original derivation of energy
current and charge current \cite{RK}, the transmission probability
in the 1D channel was ordered to $1$; meanwhile, the Hawking fluxes
were derived without the consideration of the back reaction. One can
investigate the fluxes of energy and charge and the Hawking fluxes
with the consideration of the transmission probability and the back
reaction at the same times.

\section{Hawking radiation in the Kerr-Newman spacetime}

For charged particles flowing in a 1D channel, there are
not only energy flux but also charge flux. In this section,
we investigate this case by Hawking radiation of a charged particle
in the Kerr-Newman black hole \cite{KNJ}. We first focus our
attention on the Hawking fluxes of the particle. Replacing $\Delta $
in the metric (\ref{eq:9}) with $\bar {\Delta } = r^2 - 2Mr + a^2 +
Q^2 $, we get the Kerr-Newman metric, which describes a charged
rotating spacetime with the electromagnetic potential

\begin{equation}
A_\mu = A_t dt + A_\varphi d\varphi = \frac{Qr}{r^2 + a^2\cos
^2\theta }dt - \frac{Qra\sin ^2\theta }{r^2 + a^2\cos ^2\theta
}d\varphi . \label{eq:19}
\end{equation}

\noindent The angular velocity $\Omega_H$ and the Hawking
temperature $T_H$ at the outer horizon can be easily derived and
have the same expressions as eq. (\ref{eq:11}) with the different
locations of the outer (inner) horizons

\begin{equation}
r_H = M + \sqrt {M^2 - Q^2 - a^2} , \quad r_ - = M - \sqrt {M^2 -
Q^2 - a^2} . \label{eq:20}
\end{equation}

\noindent There are much work on the thermodynamic property of the
Kerr-Newman black hole, and the radiation spectrum of a charged
particle with energy $\omega$, charge $q$ and magnetic quantum
number $m$ in the Kerr-Newman spacetime was derived \cite{ZGL} as

\begin{equation}
N_{q,m}^\pm \left(\omega \right) = \frac{1}{e^{\frac{\omega  -
m\Omega_H - q\Phi_H}{k_B T_H}}\pm 1}, \label{eq:21}
\end{equation}

\noindent where $\Phi_H = \frac{Qr_H}{r_H^2 + a^2}$ is the
electromagnetic potential at the outer horizon. Due to the existence
of the electromagnetic field and dragging field in the black hole,
the Hawking fluxes of a charged particle are consisted of three
parts: the charge flux, the angular momentum flux and the
energy-momentum tensor flux. We first derive the Hawking fluxes of
the boson. By using eq. (\ref{eq:21}), we can easily derive the
fluxes of charge, angular momentum and energy-momentum tensor as

\begin{equation}
F_q = q\int\limits_0^\infty {\frac{N_{q,m}^ - \left( \omega
\right)}{2\pi \hbar }d\omega } = \frac{q}{2\pi \hbar
}\int\limits_0^\infty {\frac{1}{e^{\frac{\omega - m\Omega _H
-q\Phi_H}{k_B T_H }} - 1}d\omega }=\frac{q k_B T_{H} }{2\pi \hbar }\int\limits_{ - \frac{ m\Omega_H +
q\Phi _H}{k_B T_{H} }}^\infty {\frac{1}{e^x - 1}dx} ,  \label{eq:22}
\end{equation}

\begin{equation}
F_a = m\int\limits_0^\infty {\frac{N_{q,m}^ - \left( \omega
\right)}{2\pi \hbar }d\omega } = \frac{m}{2\pi \hbar
}\int\limits_0^\infty {\frac{1}{e^{\frac{\omega - m\Omega _H
-q\Phi_H}{k_B T_H }} - 1}d\omega }=\frac{m k_B T_{H} }{2\pi \hbar }\int\limits_{ - \frac{ m\Omega_H +
q\Phi _H}{k_B T_{H} }}^\infty {\frac{1}{e^x - 1}dx} ,  \label{eq:23}
\end{equation}

\begin{equation}
F_M = \int\limits_0^\infty {\frac{N_m^- \left( \omega \right)}{2\pi
\hbar }\omega d\omega } = \frac{1}{2\pi \hbar }\int\limits_0^\infty
{\frac{\omega }{e^{\frac{\omega - m\Omega _H -q\Phi_H}{k_B T_H }} -
1}d\omega }= \frac{\left( {k_B T_{H} } \right)^2}{2\pi \hbar }\int\limits_{ -
\frac{ m\Omega_H + q\Phi _H }{k_B T_{H} }}^\infty {\frac{1}{e^x -
1}\left( {x + \frac{ m\Omega_H + q\Phi _H }{k_B T_{H} }} \right)dx}
.  \label{eq:24}
\end{equation}

\noindent The last equal signs in eqs. (\ref{eq:22})-(\ref{eq:24})
are obtained after we perform transformation $x = \frac{\omega - m\Omega
_H + q\Phi_H}{k_{B}T_{H}}$ . As it was explained in section 3, it is
difficult to get the integral value of the above equations because of the
particularity of the bosonic system. However, it is very easy for a
photon. For the photon, its rest mass is zero. Thus the Hawking flux
only contains the energy-momentum tensor flux, which has the same
expression as eq. (\ref{eq:14}) with the different Hawking
temperatures.

For the charged fermion, the Hawking fluxes contain the
contributions of fermions and anti-fermions \cite{Davies}. So the
charge flux, the angular momentum flux and the energy-momentum
tensor flux are obtained as

\begin{equation}
F_q = q\int\limits_0^\infty {\frac{d\omega }{2\pi \hbar }\left(
{N_{q,m} \left( \omega \right) - N_{ - q, - m} \left( \omega
\right)} \right)} = \frac{q\left( {q\Phi_H + m\Omega_H  }
\right)}{2\pi \hbar }, \label{eq:25}
\end{equation}

\begin{equation}
F_a = m\int\limits_0^\infty {\frac{d\omega }{2\pi \hbar }\left(
{N_{q,m} \left( \omega \right) - N_{ - q, - m} \left( \omega
\right)} \right)} = \frac{m\left( {q\Phi_H + m\Omega_H }
\right)}{2\pi \hbar }, \label{eq:26}
\end{equation}

\begin{equation}
F_M = \int\limits_0^\infty {\frac{\omega d\omega }{2\pi \hbar
}\left( {N_{q,m} \left( \omega \right) + N_{ - q, - m} \left( \omega
\right)} \right)} = \frac{\left( {q\Phi_H + m\Omega_H  }
\right)^2}{4\pi \hbar } + \frac{\pi k_B^2 T_H^2}{12\hbar }.
\label{eq:27}
\end{equation}

\noindent Now we compare the Hawking fluxes with the charge current
and energy current in the 1D channel. In the 1D channel mode, if
there are several kinds of charge in the reservoirs, then the
transmissions of every type of charge will produce the corresponding
fluxes. For the charged fermion, thus the total charge currents and
energy current are expressed as

\begin{equation}
F_{q_i} = I_{iR} - I_{iL} = \frac{q_i}{2\pi \hbar}(\mu_{iR} -
\mu_{iL} ), \label{eq:28}
\end{equation}

\begin{equation}
\dot {E} = \dot {U}_R - \dot {U}_L = \frac{1}{4\pi \hbar}(\mu_{iR}
^2 - \mu_{iL} ^2 ) + \frac{\pi k_B^2}{12\hbar }(T_R^2 - T_L^2  ),
\label{eq:29}
\end{equation}

\noindent where $i=1, 2, 3 \cdots$, $q_i$ and $\mu_i$ denote the
different charges and the corresponding chemical potential,
respectively.

When a charged particle is emitted in the Kerr-Newman black hole,
its motion is affected by both of the electromagnetical field and
the dragging field. So the chemical potential of the charged particle in the
black hole should reflect the effects of the electromagnetical
field and the dragging field. As it was explained in section 3, the
magnetic quantum number is regarded as a general/gauge charge in
this paper. Therefore the chemical potential is that the
electromagnetical potential plus the general/gauge potential
corresponded to the magnetic quantum number, namely $\mu_i = q\Phi_H
+ m\Omega_H$. The outside environment surrounding the black hole is
seen as $T_E = 0$ \cite{NBN} and the corresponding chemical
potential of the charged particle in this environment is
$\mu_{iE}=0$. Let $T_R = T_H$, $T_L = T_E $ and $\mu_{iR} = \mu_i$,
$\mu_{iL} = \mu_{iE}$, we find the total energy flux (\ref{eq:29})
is identical to the energy-momentum tensor flux (\ref{eq:27}).
Meanwhile, when $q_i$ respectively denotes the charge $q$ and the
magnetic quantum number $m$, we can also find that the charge fluxes
(\ref{eq:28}) are respectively identical to the charge flux
(\ref{eq:25}) and the angular momentum flux (\ref{eq:26}). This
shows the Hawking radiation of the charged fermion in the
Kerr-Newman spacetime can be described by the 1D channel. For the
charged boson, if let $T_R = T_H$, $T_L = T_E $ and $\mu_{iR} =
\mu_i$, $\mu_{iL} = \mu_{iE}$, we can also get that the charge
currents and energy current in the 1D channel obtained from eqs. (4)
and (5) are identical to the charge flux (22), angular momentum flux
(23) and energy-momentum tensor flux (24), respectively. This shows
the Hawking radiation of the boson in the Kerr-Newman spacetime can
be described by the 1D channel.

When an observer is in the dragging field, he/she moves with the
dragging field and can not fell feel the flow of the angular
momentum flux, which has been explained in section 3. In the
Kerr-Newman spcetime, therefore, he/she only feels the flows of the
charge flux and the energy-momentum flux. This case is not discussed
in this paper.

If the black hole is seen as 1D \cite{BM}, we can use parameters of
the 1D channel to describe that of the black holes. In section 3,
the thermal conductance of the 1D channel has been derived. We use
it to describe that of the Kerr-Newman black hole. Thus the thermal
conductance of this black hole is $\kappa = \frac{\pi k_B^2
T_H}{6\hbar }$, which is only related to the Hawking temperature.
Now we go on the investigation of the electric conductance. In
\cite{IL,RK}, the electric conductance was derived as $G =
\frac{q^2}{2\pi \hbar}$, which is only related to the charge of the
particle.

\section{Discussions and Conclusions}

In this paper, we have investigated the fluxes of energy and charge
in the 1D channel and Hawking radiation in the Kerr and Kerr-Newman
black holes. It shown that Hawking radiation of the bosons and the
fermions in the Kerr and Kerr-Newman black holes can be described by
the 1D channel. Our result is in consistence with that of Nation
\cite{NBN} and gives the proof of the view of Bekenstein and Mayo \cite{BM}.
In this paper, the magnetic quantum number of the
emission particle was regarded as a general charge, so the
corresponding flux was obtained and identical to the charge flux in
the 1D channel. In recent work \cite{RK,KM,BV,NBN}, there is some
work on the energy flux and entropic flux of particles in the 1D
channel. However, considering the particularity of the boson (except
for photons), they have not derived values of the flux of energy and
entropy. This was shown in the second section. In this paper, we
also didn't get the values, but this have not affected our final
result.

If black holes can be seen as 1D, we can use the thermal conductance
of the 1D to describe that of the Kerr and the Kerr-Newman black
holes. They were derived by the 1D channel and only related to the
Hawking temperatures. For the Kerr-Newman black hole, we got the
electric conductance, which is related to the charge of the
particle. In this paper, the transmission probability in
the channel was ordered to $1$, and the Hawking fluxes were
derived without considering the back reaction. So it is meaningful
to investigate the energy flux and the Hawking fluxes with
consideration of the transmission probability and the back reaction
at the same times.

\bigskip
\noindent
{\bf Acknowledgements}
This work is supported by Fundamental Research Funds for the
Central Universities (Grant No. ZYGX2009X008),  NSFC (Grant
No.10705008) and NCET. When this manuscript was finished, we found
\cite{ZZL} appeared, in which the Hawking radiation was also
researched by the one-dimensional quantum channel.


\bigskip

\begin{thebibliography}{99}

\small

\bibitem{Hawking}
S.~W.~Hawking, Commun. Math. Phys. {\bf43} 199 (1975).

\bibitem{DR}
T.~Damour and R.~Ruffini, Phys.\ Rev.\ D {\bf 14} 332 (1976).

\bibitem{SS}
S.~Sannan, Gen. Rel. Grav. {\bf20} 239 (1988).

\bibitem{KWP}
P.~Kraus and F.~Wilczek, Nucl. Phys. B {\bf437} 231 (1995);
\emph{ibid.} {\bf433} (1995) 403; M.~K.~Parikh and F.~Wilczek, Phys.
Rev. Lett. {\bf85} 5042 (2000).

\bibitem{WU}
W. Unruh, Phys. Rev. D {\bf14} 870 (1976).

\bibitem{BMPS}
R.~Brout, S.~Masser, R.~Parentani and P.~Spindel, Phys. Rep. 260,
329 (1995).

\bibitem{TP}
T.~Padmanabhan, Phys. Rept. {\bf406} 49 (2005).

\bibitem{BZ}
W.~A.~Bardeen and B.~Zumino, Nucl. Phys. B {\bf244} 421 (1984).

\bibitem{CFV}
S.~Christensen and S.~Fulling, Phys. Rev. D {\bf15} 2088 (1977);
E.~C.~Vagenas, Phys. Rev. D {\bf64} 124022 (2001).

\bibitem{RW}
S.~P.~Robinson and F.~Wilczek, Phys. Rev. Lett. {\bf95} 011303
(2005).

\bibitem{IUW1}
S.~Isoa, H.~Umetsub and F.~Wilczek, Phys. Rev. Lett. {\bf96} 151302
(2006).

\bibitem{Maldacena}
J.~M.~Maldacena, Adv. Theor. Math. Phys. {\bf2} 231 (1998);

\bibitem{Susskind}
L.~Susskind, J. Math. Phys. {\bf36} 6377 (1995).

\bibitem{BM}
J.~D.~Bekenstein and A.~E.~Mayo, Gen. Rel. Grav. {\bf33} 2095
(2001).

\bibitem{JBP}
J.~B.~Pendry, J. Phys. A {\bf16} 2161 (1983).

\bibitem{IL}
Y.~Imry and R.~Landauer, Rev. Mod. Phys. {\bf71} 306 (1999).

\bibitem{RK}
L.~G.~C.~Rego and G.~Kirczenow, Phys. Rev. B {\bf59} 13080 (1999).

\bibitem{KM}
I.~V.~Krive and E.~R.~Mucciolo, Phys. Rev. B {\bf60} 1429 (1999);

\bibitem{BV}
M.~P.~Blencowe and V.~Vitelli, Phys. Rev. A {\bf62} 052104 (2000);
M.~P.~Blencowe, Phys. Rep. {\bf395} 159 (2004); K. Schwab et al.,
Nature {\bf404} 974 (2000).

\bibitem{SUNBRB}
R.~Schutzhold1 and W.~G.~Unruh, Phys. Rev. Lett. {\bf95} 031301
(2005); P.~D.~Nation, M.~P.~Blencowe, A.~J.~Rimberg and E.~Buks,
Phys. Rev. Lett. {\bf103} 087004 (2009).

\bibitem{NBN}
P.~D.~Nation, M.~P.~Blencowe and F.~Nori,``Landauer Transport Model
for Hawking Radiation from a Black Hole'', arXiv:1009.3974 [gr-qc].

\bibitem{YSW}
Y.~S.~Wu, Phys. Rev. Lett. {\bf73} 922 (1994).

\bibitem{Kerr}
R.~P.~Kerr, Phys.\ Rev.\ Lett. {\bf 11} 237 (1963).

\bibitem{IUW2}
S.~Isoa, H.~Umetsub and F.~Wilczek, Phys. Rev. D {\bf74} 044017,
(2006).

\bibitem{DU}
P.~C.~W.~avies and W.~G.~Unruh, Proc. R. Soc. Lond. A. {\bf356} 259
(1977).

\bibitem{Davies}
P.~C.~W.~Davies, J. Phys. A: Math. Gen. {\bf11} 179 (1978).

\bibitem{LX}
L.~Liu and D.~Y.~Xu, Acta Phys. Sinica  {\bf29} 1617 (1980).
(Chinese)

\bibitem{KNJ}
E.~T.~Newman and A.~I.~Janis, J.\ Math.\ Phys. (N.Y.) {\bf6} 915
(1965).

\bibitem{JW}
Q.~Q.~Jiang and S.~Q.~Wu, Phys. Lett. B {\bf647} 200 (2007).

\bibitem{ZGL}
Z.~Zhao, Y.~X.~Gui and L.~Liu, J. Astrophysics {\bf1} 141
(1981)(Chinese).

\bibitem{ZZL}
S.~W.~Zhou, X.~X.~Zeng and W.~B.~Liu, "Hawking radiation from a BTZ
black hole viewed as Landauer transport", arXiv:1106.0559[gr-qc];
S.~W.~Zhou, X.~X.~Zeng and W.~B.~Liu,"Landauer transport model for
Hawking radiation from a Reissner-Nordstrom black hole",
arXiv:1106.0549[gr-qc]; S.~W.~Zhou and W.~B.~Liu, "Hawking radiation
via Landauer transport model", arXiv:1106.0548[gr-qc]



\end{thebibliography}
\end{document}